\documentclass [12pt,english]{article}
\usepackage{amssymb}
\usepackage{amsmath}
\usepackage{amsfonts}
\usepackage[english]{babel}

\usepackage[T2A]{fontenc}
\usepackage[cp1251]{inputenc}
\usepackage{fullpage}
\usepackage{bm}
\usepackage{graphicx}
\numberwithin{equation}{section}
\newcounter{Alpha}

\begin{document}

\title{\LARGE 
Localized excitons in 1D--dimensional half--filled paramagnetic  Hubbard model.}

\author{N. I. Chashchin\thanks{E--mail: nik.iv.chaschin@mail.ru} \\ \textit{Ural State Forestry University}\\
%, Russian Academy of Sciences}
\textit {Ekaterinburg, Sibirskii trakt 37, 620100 Russia}}
\date{}

\maketitle

\begin{abstract}
By the example of the Hubbard model
we analytically and numerically examine the formating and coexisting of localized electron--electron pairs (doublons) and localized electron--hole pairs (Frenkel--type excitons) . Here we demonstrate that at a variation of the on-site  Coulomb repulsion U there occurs a quantum transition from the doublon to the exciton regime that is conditioned by the low-energy effective hybridization of the valence and conductivity subbands.
We calculate momentum distribution functions and  electronic spectrum functions for different U, which reveal topologically nontrivial behaviour at the Fermi level. 
\end{abstract}

{\bf Keywords:} Hubbard model, electronic spectrum function, exciton,
doubly occupied sites, momentum distribution function, nontrivial topology.
\\
{}
\\

\section{\!\!\!\!\!\!. Introduction}
Interplay of band theory and strong electron correlation effects has been a basic phenomenon for understanding
physics of many body models and materials. Electron repulsion due to local Coulomb interactions tends to localize electrons, on the other hand, kinetic effects are favourable to electron
itineracy. The energetic competition between them lead to different phases of comparable magnitude.

The single band Hubbard model (HM) \cite{Hubbard1964} with one electron per site (half--filled) is the simplest model and prototypical example for investigation of  correlated electrons.  
The initial band of noninteracting electrons is splitted into two subbands in correlated state: the lower -- valence band and the upper -- conductivity band. The electronic spectrum of a crystal can be organized in the form of energy bands as a function of the crystal momentum k, which is guaranteed to be a good quantum number due to the translational symmetry of the lattice. When an electron is excited from the valence into the conductivity band it leaves behind 
a vacancy -- a hole. %The hole has positive charge equal to the electron's charge. 
In a divalent crystal \cite{Kozlov1965} electrons are concentrated near minimums of the conductivity band and holes  near maximums of the valence band. 
Nearby these extremums band energy spectra in simple form are $E(k)=\hbar^2 k^2/2 m^*$, where $m^*$ is an effective mass of the electrons or the holes correspondingly. The vacant place (hole)  in the filled valence band has the negative mass in definition, though it is convenient to consider this hole as a real particle with the positive mass
and charge. 

The double occupancy is a correlation function of a local pair of electrons with antiparallel spins 
$\langle n_{\uparrow}n_{\downarrow}\rangle$.  
The parameter plays an important role in the study of correlated systems, the Mott metal--insulator transition 
\cite{Mott1990}, the semiconductor - semimetal mixing \cite{Bronold2006}, quantum phase transition \cite{Irkhin2019}, local moment formation, transport properties \cite{Rontani2005}. The local Coulomb repulsion  disfavors doubly-occupied sites, and decreases the double occupancy \cite{Tudor2001, Erik2016}. 
%Given this important role, it is worthwhile to study the matter more carefully.

It is appropriate to assume that there also has to be   
a local bound state of an electron in the conductivity band and a hole in the valence band which are attracted to each other by the Coulomb force. The localazed on a site Frenkel--type exciton $\langle(1-n_{\sigma})n_{\bar\sigma}\rangle$ is an electrically neutral quasiparticle that exists in various electronic phases of the model \cite{Plakida2011, Zenker2014, Mazziotti2020}. 
%insulators, semiconductors, semimetals, and excitonic insulators. 
The exciton is regarded, as an elementary excitation of condensed matter, which can transport energy without transporting electric charge.
The correlation effects caused by these bosonic--type single quantum states are of an interest in the study of electronic systems.

\section{\!\!\!\!\!\!. Theoretical Model and Method}

Consider the Hubbard model in the representation  we got in our previous works 
\cite{chaschin2011_1,chaschin2011_2,chaschin2012_3,chaschin2016_4} by using the method of generating functional of Green functions with the subsequent Legendre transformation. 
The Hamiltonian of the half-filled and symmetrical Hubbard model is 
\begin{equation}
\mathcal{H}= -t\sum\limits_{\langle i,j\rangle\sigma}c_{i\sigma}^{\dag}c_{j\sigma}+\sum\limits_{i\sigma}
\varepsilon_\sigma n_{i\sigma}
+U\sum\limits_{i}n_{i\uparrow}n_{j\downarrow}\,,
\label{Pm:Hub_ham}
\end{equation} 
where U is the parameter of the Coulomb interaction
at a site; $c_{i\sigma} (c_{i\sigma}^\dag)$ are fermion operators that describe
the annihilation (generation) of electrons with spins
up and down ($\sigma=\uparrow,\downarrow$); $n_{i\sigma}$ indicates the operators of
the number of particles; $t$ is the parameter of hopping of electrons from site to site; and in the designation 
$\langle i,j\rangle$  crystall sites  are the nearest, $\varepsilon_\sigma=-\displaystyle{\sigma\frac{h}{2}-\mu}$, 
where $h=g\mu_BH$, $g$ is the electron $g$--factor, $\mu_B$ is the Bohr magneton, $H$ is the external magnetic field, and $\mu$ is the chemical potential. 

The general equations derived in \cite{chaschin2016_4} are valid for different solutions --- paramagnetic or magnetic. Restricting ourselves to the paramagnetic case, we get connected systems of  equations for two types of Green functions. 

The fermionic charge propagator ($N=2\,G$)
%G_\uparrow+G_\downarrow=  
\begin{equation}
\displaystyle N({\bf k}, i\omega_n)=\frac{2}{i\omega_n-\varepsilon_k-\Sigma({\bf k},i\omega_n)}\,,
\label{N_el}
\end{equation} 
where $\varepsilon_k=-\cos(k)$ is the 1D electronic free spectrum, 
$\omega_n=(2n+1)\, \pi T$ ($n=0,\pm 1,\pm 2,\dots$) are fermionic Matsubara frequencies, and 
$\Sigma({\bf k},i\omega_)$ is the fermionic particle self energy. 

The bosonic propagator 

\begin{equation}
\displaystyle Q({\bf q},i\Omega_\nu)=-\frac{1}{1+\frac{U}{2}\Pi(q,i\Omega_\nu)}\,,
\label{Q_bos}
\end{equation}
where $\Omega_\nu=2\, \nu \,\pi T$ ($\nu=0,\pm 1,\pm 2,\dots$) are bosonic Matsubara frequencies, and 
$\Pi({\bf q},i\Omega_\nu)$ is the bosonic particle self energy.

The Green function of electrons (\ref{N_el}) lets the representation of the charge propagator as 
$N = G_{-}+\,G_{+}$, and the band-to-band propagator $N^{\,'}=G_{-}-\,G_{+}$, where the indexes $\pm $ correspond to  separation of the initial electron band into the valence and conductivity subbands. 

\begin{equation}
\begin{array}{l}
\displaystyle G_{-}(k, i\omega_n) = \frac{1}{i\,\omega_n-\varepsilon_k-
\displaystyle \Sigma_{-}(k, i\omega_n)}\,,
\\{}\\ 
\displaystyle G_{+}(k, i\omega_n) = \frac{1}{i\,\omega_n+\varepsilon_k-
\displaystyle\Sigma_{+}(k, i\omega_n)}\,;
\end{array}
\label{G_fin} 
\end{equation}

The electronic self energy part is evaluated by the next set of equations 

\begin{equation}
\left\{
\begin{array}{l}
\displaystyle \Im \Sigma_+(k,\omega) = \frac{U}{4}\sum\limits_{q^\prime}
\displaystyle \left[\left(1-\tanh(\frac{\varepsilon_{q^\prime}}{2 T})
\displaystyle \tanh(\frac{\varepsilon_{q^\prime}-\omega}{2 T})\right)\right.
\displaystyle\Im Q_-(q^\prime-k ;\varepsilon_{q^\prime}-\omega)+\\ 
\hspace{4cm}
\displaystyle  \left.\left(1-\tanh(\frac{\varepsilon_{q^\prime}}{2 T})
\displaystyle \tanh(\frac{\varepsilon_{k^\prime}+\omega}{2 T})\right)
\displaystyle\Im Q_+(q^\prime-k;\,-\varepsilon_{k^\prime}-\omega)\right]\,,
\\
{}
\\
\displaystyle \Re \Sigma_+(k,\omega) = \frac{1}{\pi}\int_{-\infty}^\infty  
\displaystyle \frac{\Im \Sigma_+(k,\omega^\prime)}{\omega^\prime - \omega}\,
d\omega^\prime\,;
\\
{}
\\
\displaystyle \Im\Sigma_+(k+\pi,\omega) = \Im\Sigma_-(k,\omega), \quad \Re\Sigma_+(k+\pi,\omega) = \Re\Sigma_-(k,\omega)\,;
\\
{}
\\
\displaystyle \Im\Sigma_+(k,-\omega) = \Im\Sigma_-(k,\omega), \quad \Re\Sigma_+(k,-\omega) = 
-\,\Re\Sigma_-(k,\omega)\,.
\end{array}
\right.
\label{Sigma}
\end{equation}

The bosonic self energy part is evaluated by the equations 

\begin{equation}
\left\{
\begin{array}{l}
\displaystyle \Im\Pi_+(q,\Omega) = \frac{U}{4}\sum\limits_{k^\prime}
\displaystyle \left[\left(1-\tanh(\frac{\varepsilon_{k^\prime}}{2 T})
\displaystyle \tanh(\frac{\varepsilon_{k^\prime}-\Omega}{2 T})\right)\right.
\Im G_-(k^\prime-q;\varepsilon_{k^\prime}-\Omega)+\\
\hspace{3.3cm}
\displaystyle  \left.\left(1-\tanh(\frac{\varepsilon_{k^\prime}}{2 T})
\displaystyle \tanh(\frac{\varepsilon_{k^\prime}+\Omega}{2 T})\right)
\Im G_+(k^\prime-q;-\varepsilon_{k^\prime}-\Omega)\right],
\\
{}
\\
\displaystyle \Re \Pi_+(q,\Omega) = \frac{1}{\pi}\int_{-\infty}^\infty  
\displaystyle \frac{\tanh(\frac{\Omega^\prime}{2 T})\Im \Pi_+(q,\Omega^\prime)}{\Omega^\prime - \Omega}\,
d\Omega^\prime\,; 
\\
{}
\\
\Im Q_+(q;\Omega)=\displaystyle\frac
{\,\Im \Pi_+(q;\Omega)}{\left[1+\Re\Pi_+(q;\Omega)\right]^2+
\left[\Im \Pi_+(q;\Omega)\tanh(\frac{\Omega}{2 T})\right]^2}\,; 
\\
{}
\\
\displaystyle\Im\Pi_{+}(k+\pi,\omega) = \Im\Pi_{-}(k,\omega),\quad  \Re \Pi_{+}(k+\pi,\omega) = 
\Re \Pi_{-}(k,\omega)\,; 
\\
{}
\\
\displaystyle \Im\Pi_{+}(k,-\omega) = -\,\Im\Pi_{-}(k,\omega),\quad  \Re \Pi(k,-\omega) = 
\Re \Pi_{-}(k,\omega)\,. 
\end{array}
\right.
\label{PI_Q}
\end{equation}%Yet note that all temperature dependences of the model is confined by a unique factor 

Thus, we get two coupled sets of the integral equations (\ref{Sigma}, \ref{PI_Q}) which can be numerically calculated and investigated.   
As we see, they are structurally identical over the mutual substitutions: ${\Im N}\leftrightarrow{\Im Q}$, ${\Im\Sigma}\leftrightarrow{\Im\Pi}$, and ${\Re\Sigma}\leftrightarrow{\Re\Pi}$.  

The imaginary and real parts of the Green functions obey to the next symmetry properties: 
\begin{equation}
\begin{array}{l}
\displaystyle \Im G_+(k \pm \pi,\omega) = \Im G_-(k,\omega),\quad \Re G_+(k \pm \pi,\omega) = 
\Re G_-(k,\omega); 
\\
\displaystyle \Im G_+(k,-\omega) = \Im G_-(k,\omega),\quad \Re G_+(k,-\omega) = 
-\:\Re G_-(k,\omega)\,.
\end{array}
\label{Pm:G_sym}
\end{equation}

\section{\!\!\!\!\!\!. Results and Discussion}

\addtocounter{Alpha}{1}
{\bf\large{\Alph{Alpha}. Double occupancy states and local excitons}}\\ %\setcounter{equation}{1}
\setcounter{Alpha}{1}
\addtocounter{Alpha}{1}

Write down the two--particle Green function in the explicit form 
\begin{equation}
\displaystyle \langle \hat T\,n_{1\uparrow}n_{1\downarrow}\rangle = 
\displaystyle \langle (n_{1\uparrow}-\frac{1}{2})(n_{1\downarrow}-\frac{1}{2})\rangle = 
\langle n_{1\uparrow}n_{1\downarrow}\rangle -\frac{\langle n_1\rangle}{2}+\frac{1}{4}\,,
%\tag{A.1}
\label{A_Dbl}
\end{equation}
as in our symmetrical representation $\displaystyle\hat T n_{1\sigma}=n_{1\sigma}-\frac{1}{2}$. 

From the other hand \cite{chaschin2011_1}, 

\begin{equation}
%\begin{array}{l}
\displaystyle \langle \hat T\,n_{1\uparrow}n_{1\downarrow}\rangle = 
\frac{1}{4}\left[\frac{\delta^2\Phi}{\delta \rho(11)\,{\delta\rho(11)}}-
\frac{\delta^2\Phi}{\delta \eta(11)\,{\delta\eta(11)}}\right]+ 
\frac{1}{4}\left[\left(\frac{\delta\Phi}{\delta \rho(11)}\right)^2 - 
\left(\frac{\delta\Phi}{\delta \eta(11)}\right)^2\right]\,,
%\end{array}
\label{A_Dbl_GF}   
\end{equation}

where $\displaystyle\frac{\delta\Phi}{\delta \rho(11)}=\langle n_1\rangle-1$, 
$\displaystyle\frac{\delta\Phi}{\delta \eta(11)}=\langle n_{1\uparrow}-n_{1\downarrow}\rangle$;
$\Phi=\ln Z$ is the generated functional of the connected Green functions; $1\equiv (i,\tau)$, i--a crystal site, 
$\tau$--an imaginary time; $\eta(12)=h\,\delta_{12}$ is the external magnetic field; 
$M_1=\langle n_{1\uparrow}-n_{1\downarrow}\rangle$ --- the local magnetic moment.
%, and $\hat T$ is the imaginary time ordering operator in the definition of Green functions; $\hat T %n_{1\sigma}=n_{1\sigma}-1/2$, $\hat T M_{1}=M_{1}$ in the sym--form representation. 

As we see from (\ref{A_Dbl_GF}), determinal factors for the propagator of double occupied sites are two--particle Green functions:   
\begin{equation}
\displaystyle \langle \hat T\, M_1^2\rangle=\frac{\delta^2\Phi}{\delta \eta(11)\,{\delta\eta(11)}},\;\;
\displaystyle \langle \hat T\, n_1^2\rangle=\frac{\delta^2\Phi}{\delta \rho(11)\,{\delta\rho(11)}}\,.
\label{DPM:XM}   
\end{equation}

Pauli exclusion principle in the functional form \cite{chaschin2011_1} is  
\[
\displaystyle\frac{\delta^2\Phi}{\delta \rho(11)\,{\delta\rho(11)}}+
\frac{\delta^2\Phi}{\delta \eta(11)\,{\delta\eta(11)}}+
\left(\frac{\delta\Phi}{\delta \rho(11)}\right)^2+
\left(\frac{\delta\Phi}{\delta \eta(11)}\right)^2 = 0,
\]  

which allow to write (\ref{A_Dbl_GF}) in the form 
\begin{equation}
\displaystyle\langle \hat T\,n_{1\uparrow}n_{1\downarrow}\rangle =
-\frac{1}{2}\,\frac{\delta^2\Phi}{\delta \eta(11)\,{\delta\eta(11)}}-
\frac{1}{2}(\langle M_1\rangle)^2\,.
\label{A_Dbl_2}
\end{equation}
Comparing two expressions (\ref{A_Dbl}, \ref{A_Dbl_2}) we get general formula for the number of double--occupied sites 

\begin{equation}
\displaystyle\langle n_{1\uparrow}n_{1\downarrow}\rangle = \frac{\langle n_1\rangle}{2}
-\frac{1}{4}-\frac{(\langle M_1\rangle)^2}{2}- 
\frac{\langle\hat T\, M^2_1\rangle}{2}\,.
%-\frac{1}{2}\,\frac{\delta^2\Phi}{\delta \eta(11)\,{\delta\eta(11)}}\,.
\label{A_Dbl_M}
\end{equation}

%The momentum distribution function $n(k)$ per one spin direction gives an average occupation number of electronic %states of the momentum $k$   
The mathematical expression for $\langle\hat T\, M^2\rangle$ was earlier derived in \cite{chaschin2016_4}:

\begin{equation}
\begin{array}{l}
\displaystyle \langle\hat T\, M^2\rangle =
-\frac{1}{\pi U}\sum\limits_{k}\int_{-\infty}^\infty 
\tanh\left(\frac{\omega}{2 T}\right)\left(\omega-\varepsilon(k)\right)\,\Im G_-(k,\omega)\,d\omega\,= 
\\
\\
\displaystyle -\frac{1}{\pi^2 U}\int_{-1}^{1} \int_{-\infty}^\infty \frac{d\varepsilon}{\sqrt{1-\varepsilon^2}}
\tanh\left(\frac{\omega}{2 T}\right)
\left(\omega-\varepsilon\right)\,\Im G_-(\varepsilon,\omega)\,d\omega\,.
\end{array}
\label{Pm:MM}
\end{equation}

Finally for paramagnetic $\langle M\rangle=0$ and half--filled $\langle n\rangle=1$ solution the average number of double--occupied sites (\ref{A_Dbl_M}) is  
\begin{equation}
\displaystyle \langle n^e_{\uparrow}n^e_{\downarrow}\rangle=\frac{1}{4}-\frac{\langle\hat T\, M^2\rangle}{2} 
\equiv\mathcal{X}(U)\,,   
\label{PM:Dbl}   
\end{equation}
where we denote now $n^e_{\sigma}$ as the operator for electrons. 

Put the next graphic designations:
\unitlength=1mm \,
\begin{picture}(4,6)
\put(2,1){\circle{3}}
\end{picture} is just a vacant place uncorrelated with other fermions, and unbound to a host--site; 
\begin{picture}(4,6)
\put(2,1){\circle{3}}
\put(.5,1){\line(1,0){3}}
\put(2,2.5){\vector(0,1){3}}
\end{picture} is a negative charged electron; 
\begin{picture}(4,6)
\put(2,1){\circle{3}}
\put(2,2.5){\vector(0,1){3}}
\put(.5,1){\line(1,0){3}}
\put(2,-.5){\line(0,1){3}}
\end{picture} is a positive charged hole.
 
In that case all electronic configurations on a single site can be represented in one of particular form 
\unitlength=1mm \,
\begin{picture}(57,10)
\put(0,1){\circle{3}}
\put(3.2,1){\circle{3}}
\put(6,0){,}
\put(10,1){\circle{3}}
\put(13,1){\circle{3}}
\put(10,2.5){\vector(0,1){3}}
\put(13,-.5){\vector(0,-1){3}}
\put(8.5,1){\line(1,0){3}}
\put(11.5,1){\line(1,0){3}}
\put(16,0){,}
\put(20,1){\circle{3}}
\put(23.2,1){\circle{3}}
\put(20,2.5){\vector(0,1){3}}
\put(18.5,1){\line(1,0){3}}
\put(26,0){,}
\put(30,1){\circle{3}}
\put(33.2,1){\circle{3}}
\put(33.2,-.5){\vector(0,-1){3}}
\put(31.5,1){\line(1,0){3}}
\put(33.2,-.5){\line(0,1){3}}
\put(36,0){,}
\put(40,1){\circle{3}}
\put(43.2,1){\circle{3}}
\put(39.8,2.5){\vector(0,1){3}}
\put(43.2,-.5){\vector(0,-1){3}}
\put(38.5,1){\line(1,0){3}}
\put(41.5,1){\line(1,0){3}}
\put(43.2,-.5){\line(0,1){3}}
\put(46,0){,}
\put(50,1){\circle{3}}
\put(53.2,1){\circle{3}}
\put(50,2.5){\vector(0,1){3}}
\put(53.2,-.5){\vector(0,-1){3}}
\put(48.5,1){\line(1,0){3}}
\put(51.5,1){\line(1,0){3}}
\put(50,-.5){\line(0,1){3}}
\put(53.2,-.5){\line(0,1){3}}
\put(56,0){.}
\end{picture}\,\,

Using the operator expression for a hole's number $n^h_{\sigma}$= $1-n^e_{\sigma} $ we can write the   
obvious identity for the average number of electrons with a certain spin direction in the system as  
$\displaystyle 
\langle n^e_{\sigma}\rangle=\langle n^e_{\sigma}n^e_{\bar\sigma}\rangle +\langle n^e_{\sigma}n^h_{\bar\sigma}\rangle$; from which we get   

\begin{equation}
\begin{array}{l}
\displaystyle\langle n^e_{\sigma}n^e_{\bar\sigma}\rangle = \mathcal{X}(U)
\\ 
\displaystyle\langle n^e_{\sigma}n^h_{\bar\sigma}\rangle - \langle n^e_{\sigma}\rangle = -\mathcal{X}(U)\,.
\end{array}
\label{Dbl_Exc}
\end{equation}

\begin{figure}[h]
\begin{l}
\includegraphics[width=0.8 \textwidth]{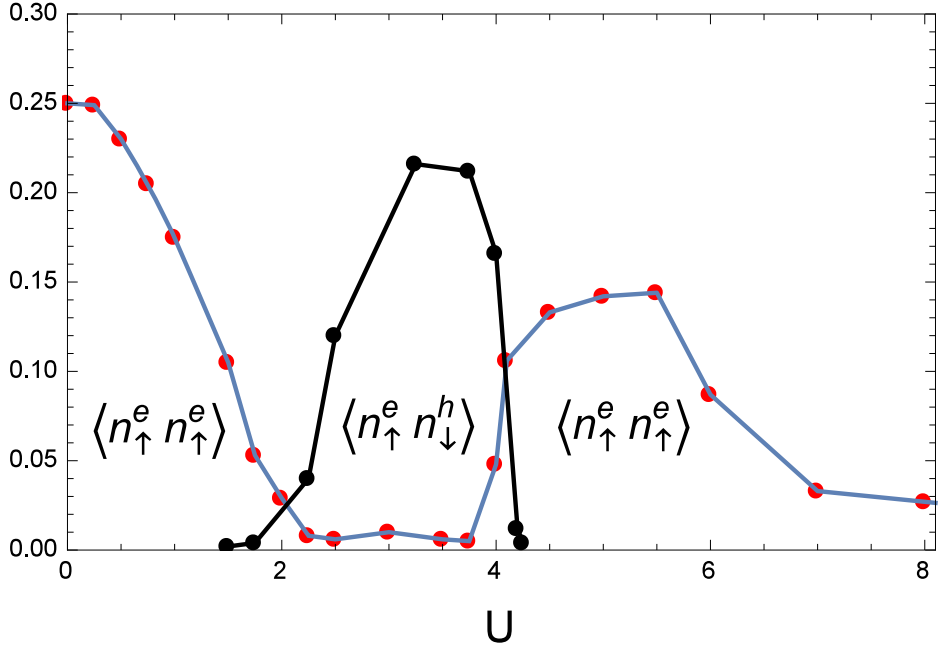}
\caption{Double--occupied and localized excitonic sites. Here 
$\langle n^e_{1\uparrow}n^e_{1\downarrow}\rangle$ is an average number of  sites with the double occupancy by electrons, 
$\langle n^e_{1\uparrow}n^h_{1\downarrow}\rangle$ is an average number of sites, which occupied by the localized excitons. The excitons are appeared and disappeared quite steeply in region $1.7<U<4.3$, but small amount of the double--occupancy also appear in the region. For another values of the Coulomb repulsion we observe only quite smoothly varying parameter of $\langle n^e_{1\uparrow}n^e_{1\downarrow}\rangle$. This situation can be considered as a sort of quantum phase transition.} 
\label{Pm:Dbl_Exc}
\end{l}
\end{figure}

A numerical value of $\langle n^e_{\uparrow}n^h_{\downarrow}\rangle$ can be expressed in the graphic view as  
$\displaystyle\langle n^e_{\uparrow}n^h_{\downarrow}\rangle$= 
$\langle\,
\begin{picture}(7,7)
\put(2,1){\circle{3}}
\put(5.2,1){\circle{3}}
\put(2,2.5){\vector(0,1){3}}
\put(5,-.5){\vector(0,-1){3}}
\put(.5,1.2){\line(1,0){3}}
\put(3.5,1.2){\line(1,0){3}}
\put(5,-.5){\line(0,1){3}}
\end{picture}\,\rangle$ + 
$\langle\,
\begin{picture}(7,7)
\put(2,1){\circle{3}}
\put(5.2,1){\circle{3}}
\put(2,2.5){\vector(0,1){3}}
\put(.5,1.2){\line(1,0){3}}
\end{picture} \,
\rangle.$
At that $\langle\,
\begin{picture}(7,7)
\put(2,1){\circle{3}}
\put(5.2,1){\circle{3}}
\put(2,2.5){\vector(0,1){3}}
\put(.5,1.2){\line(1,0){3}}
\end{picture} \,
\rangle$ = $\langle n^e_\uparrow\rangle$, and 
$\langle\,
\begin{picture}(7,7)
\put(2,1){\circle{3}}
\put(5.2,1){\circle{3}}
\put(2,2.5){\vector(0,1){3}}
\put(5,-.5){\vector(0,-1){3}}
\put(.5,1.2){\line(1,0){3}}
\put(3.5,1.2){\line(1,0){3}}
\put(5,-.5){\line(0,1){3}}
\end{picture}\,\rangle$ = $\langle\mathcal N^\text{exc}_\uparrow\rangle$ -- an excitonic pair for the single 
($\uparrow$)  spin direction. 
From the second equation of (\ref{Dbl_Exc}) we finally find the average number of localized excitons for both spin directions as 
\begin{equation}
\displaystyle\langle\mathcal N^\text{exc}\rangle = - 2\,\mathcal{X}(U)\,.
\label{N_exc}
\end{equation} 
Positive or negative value of the parameter $\mathcal{X}(U)$ for different U specifies the electronic 
state of the system, its quantum phase.  The interaction drives the bands from a semimetallic configuration to an excitonic insulating one. The excitonic nature of the new state is revealed itself in the deformed and mixed valence and conductivity subbands, which are indicating the hybridization of the original band and the correlated coexistence of electrons with holes (see Fig. \ref{Pm:Dbl_Exc}).
\\

\addtocounter{Alpha}{0}
{\bf\large{\Alph{Alpha}. Electronic spectra and momentum distribution function}}\\ %\setcounter{equation}{1}
\setcounter{Alpha}{0}
\addtocounter{Alpha}{0}

Correlated electronic energy spectra of the conductivity $E_{-}(k)$ and the valence $E_{+}(k)$ bands as functions of the momentum k are obtained as solutions of the following electronic dispersion equations 
($\Re G_{\pm}^{-1}(k,\omega+i0)=0$):
\begin{equation}
\begin{array}{l}
\displaystyle E_{-}(k) - \varepsilon(k) - \Re\Sigma_{-}(k,E_{-}(k)) = 0\,,
\\
\displaystyle  E_{+}(k) + \varepsilon(k) - \Re\Sigma_{+}(k,E_{+}(k)) = 0\,
\end{array}
\label{Pm:Disp}
\end{equation}

that have the following symmetry properties 
\begin{equation}
\displaystyle E_{\pm}(k)=E_{\pm}(k\pm \pi),\quad E_{\pm}(k)=E_{\mp}(k\pm \pi)\,.
\label{Pm:ESym} 
\end{equation}

We calculate the behaviour of the momentum distribution function per one spin direction for the ground state of the 1D Hubbard model as 
\begin{equation}
\displaystyle n(k)=\frac{1}{2}+\frac{1}{\pi}
\displaystyle\int_{-\infty}^\infty \tanh\left(\frac{\omega}{2 T}\right)\,\Im G_{-}(k,\omega)\, d \omega\,. 
\label{DPM:nk}
\end{equation}
The momentum distribution function $n(k)$ gives an average occupation number of electronic states of the momentum $k$ \cite{Voit1990}. 

\begin{figure}[h]
\begin{l}
\includegraphics[width=0.8 \textwidth]{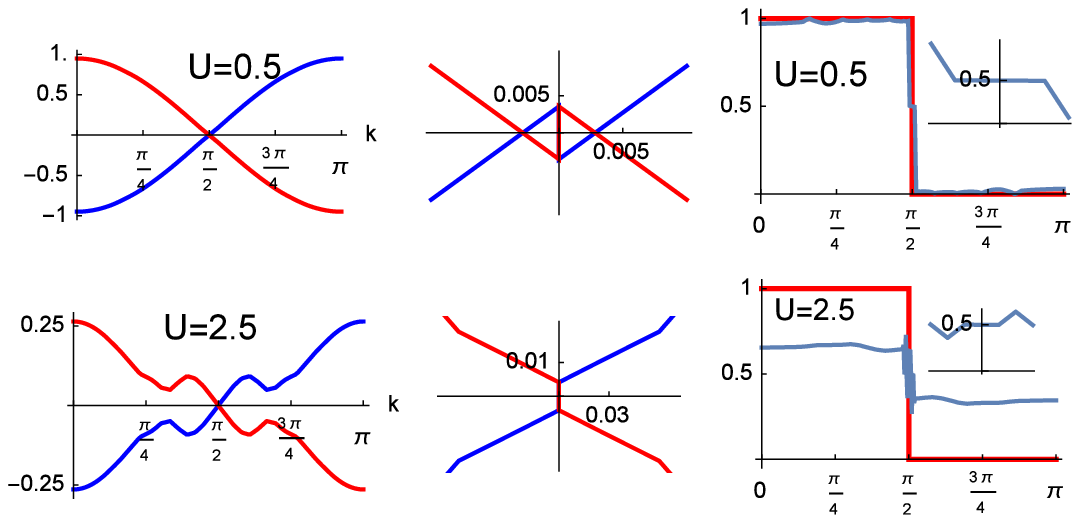}
\caption{The top and the bottom panels are corresponding to $U=0.5$} and $U=2.5$. 
The first picture in every row represents electronic spectra $E_\mp(k)$, the second one is the same but near $k_F = \pi/2$; the third picture depicts the momentum distribution functions with insets for $k\approx k_F$. A flat behaviour
of $n_k$-curve around the Fermi level demonstrates an equality of electron ($k>k_F$) and hole ($k<k_F$) numbers at the Fermi level. Note the nontrivial topology of $E_\mp(k)$ in the area under consideration. 
\label{Pm:pict}
\end{l}
\end{figure} 

A presence or absence of a finite gap at the Fermi momentum $k_F$ are generally considered for discerning
among different sorts of an electronic behavior. The finite jump would indicate that the
qusiparticle excitations are of the Fermi--liquid type without any gap in the spectrum and
therefore, the system is of a metallic--type. For other electronic phases: insulators, semiconductors, semimetls, and excitonic insulators the function's view near the $k_F$ considerably differs from the metallic--type behavior.  

The relevance of function (\ref{DPM:nk}) refers to its analitical properties. Formally one of the determinal factors distingwishing electronic phases of the system is an energy gap between the conductivity and the valence bands at the Fermi momentum: $\Delta=E_-(k_F)-E_+(k_F)$ which distinguish the correlated electrons from the free ones. The finite jump when $\Delta=0$  would indicate that the
qusiparticle excitations are of the Fermi--liquid type without any gap in the spectrum and
therefore, the system is of a metallic--type. At $\Delta>0$ we have an insulator, when  $\Delta\simeq 0$ we get a gapless semiconductor \cite{Tsidilk1988}
%[I.M. Tsidilkovski: Gap less Semiconductors - a New Class of Materials (Akademie,Berlin 1988); 
%I.I. Lyapilin,I.M. Tsidilkovski: Usp. Fiz. Nauk 146,35 (1985)], 
and a semimetal or the excitonic insulator \cite{Zenker2014,{Mazziotti2020}}  for $\Delta < 0$. 

In the weak-coupling limit (small U), near the semimetal to semiconductor transition the system can become unstable against the formation of multiple excitons and this can lead to a condensation state called the excitonic insulator. 
Different values of band gap for U = 0.5 and U = 2.5 permit to refer the first case to the SM - phase, and the second 
one to EI - phase.

\end{document}